\documentclass[preprint,preprintnumbers,amsmath,amssymb]{revtex4}
\usepackage{amsmath}
\usepackage{bm}
\usepackage{graphicx}
\usepackage{subfig}
\usepackage{color}
\newcommand{\be}{\begin{equation}}
\newcommand{\ee}{\end{equation}}
\newcommand{\bea}{\begin{eqnarray}}
\newcommand{\eea}{\end{eqnarray}}
\newcommand{\bc}{\begin{center}}
\newcommand{\ec}{\end{center}}

\begin{document}

\title{Chiral symmetry restoration in RQED at finite temperature in the supercritical coupling regime}

\author{Jean B\'aez Cuevas$^1$, Alfredo Raya$^{2,3}$ and J.~C. Rojas$^1$}
\email{jbc039@alumnos.ucn.cl,alfredo.raya@umich.mx,jurojas@ucn.cl}
\affiliation{$^1$Departamento de F\'{\i}sica, Universidad Cat\'olica del Norte, Casilla 1280, Antofagasta, Chile.\\
$^2$ Instituto de F\'{\i}sica y Matem\'aticas, Universidad Michoacana de San Nicol\'as de Hidalgo. Edificio C-3, Ciudad Universitaria. Francisco J. M\'ujica s/n Col. Fel\'{\i}citas del R\'{\i}o. C.~P. 58040, Morelia, Michoac\'an, Mexico.\\
$^3$Centro de Ciencias Exactas - Universidad del Bio-Bio.\\ Avda. Andr\'es Bello 720, Casilla 447, Chill\'an, Chile.
}

\begin{abstract}
We explore the conditions for chiral symmetry breaking in reduced (or pseudo) quantum electrodynamics at finite temperature in connection with graphene and other 2D-materials with an underlying Dirac behavior of the charge carriers. By solving the corresponding Schwinger-Dyson equation in   a suitable truncation (either the non-local Nash gauge including vacuum polarization effects in the large fermion family number $n_f$ limit or the quenched rainbow approximation, in a Landau-like gauge) and neglecting wavefunction renormalization effects, 
we find the need of the {\em effective} coupling to exceed a critical value $\alpha_c$ in order for chiral symmetry to be broken, in agreement with known results from other groups. In this supercritical regime, we  add the effects of a thermal bath at temperature $T$ and find the critical values of this parameter that lead to chiral symmetry restoration. 
\end{abstract}

\maketitle


\section{Introduction}

Gauge symmetry lies at the cornerstone of modern physics~\cite{tHooft:1995wad}. Apart from gravity, fundamental interactions are described by quantum field theories in which the gauge principle describes precisely the manner in which matter fields interact through the exchange of gauge bosons. Electroweak and strong interactions, for instance, are described based on the observation that both matter and gauge fields live in a four-dimensional Poincar\'e space-time~(see, for instance, \cite{Donoghue:1992dd}). Of course, these models can be formulated in space-times of different dimensionality. The Schwinger model~\cite{Schwinger:1962tp} is the incarnation of quantum electrodynamics (QED) in (1+1)-dimensions, wheres the t'\ Hooft model~\cite{tHooft:1974pnl} is the analog of quantum chromodynamics (QCD) in the same dimensionality. 
Adding gravity to the set of fundamental interactions, in order to avoid anomalies in space-time of larger dimensionality, brane-world scenarios (see, for instance, Refs.~\cite{Maartens:2003tw,Maartens:2010ar,Brax:2003fv,Brax:2004xh,Johnson:2000ch}
 and references within) require that the matter and gauge fields of the standard model of particle physics live in the four-dimensional space-time corresponding to a brane, whereas gravity fields can propagate in extra (bulk) dimensions. This analogy has been put forward to describe the low-energy behavior of graphene and other Dirac matter systems.

Graphene has been theoretically studied for over seven decades~\cite{Wallace}. Nevertheless, the experimental isolation of graphene flakes~\cite{graphene1,graphene2,graphene3} has boosted, in addition to the search of technological applications, the interest in exploring the connection of the full family of 2D Dirac matter systems and the phenomenology of high energy physics, basically because the low-energy quasiparticle excitations in these materials are described by a 2D massless Dirac equation. The long-range Coulomb interactions in graphene are introduced via minimal coupling. Nevertheless, the electromagnetic field is not restricted to the graphene plane, and therefore a mere dimensional reduction of QED to a plane to account for these interactions is inappropriate. An alternative has been proposed in terms of a gauge theory of electromagnetic interactions where the gauge and matter fields have dynamics in different space-time dimensions. Pseudo~\cite{Marino:1992xi} or Reduced QED~\cite{Gorbar:2001qt,Teber:2012de} (we adopt the latter name and refer to the theory as RQED) is a gauge theory which for graphene allows the dynamics of electrons in (2+1)-dimensions, but the electromagnetic field is described in (3+1)-D. Then, by coupling a current defined in the plane of motion of electrons, the Lagrangian of the theory develops fractional powers of the D'\ Alambertian operator, describing interesting features as compared with ordinary QED in (3+1) and (2+1)-dimensions. Photons remain transverse on the plane, but the infrared divergence in the pole of its  propagator is softened from $1/q^2 \to 1/(2\sqrt{q^2})$~\cite{Marino:1992xi,Gorbar:2001qt,Teber:2012de}.

Perturbation theory aspects of the theory have been  widely explored by several groups~\cite{Teber:2012de, Kotikov:2013kcl, Kotikov:2013eha, Teber:2014ita, Teber:2016unz, Teber:2018goo, Teber:2018qcn, Kotikov:2018wxe} up to two-loops. Non-perturbative aspects of the theory have been already addressed (see, for instance, Refs.~\cite{Gorbar:2001qt,Kotikov:2016wrb,Kotikov:2016yrn,Ahmad:2016dsb}). In particular, the Schwinger-Dyson equation (SDE) for the fermion propagator has been explored in \cite{Gorbar:2001qt} incorporating vacuum polarization effects at the leading order of the $1/n_f$  ($n_f$ representing fermion family number, in the large $n_f$ limits) approximation. The authors of that work find that using the non-local Nash gauge, it is possible to break the chiral symmetry of the massless theory if the effective coupling of the model exceeds a critical value. This corresponds to consider the number of fermion familes $n_f$  below a critical value for a fixing the electromagnetic coupling $e^2$ and vice versa. At the critical $n_f^c$, the electromagnetic coupling diverges. The dynamically generated mass in this case follows a Miransky scaling law and the full critical line in the plane $(n_f,e^2)$ is explored in detail. On different grounds, the SDE has also been considered in~\cite{Alves:2013bna} by quenching the theory truncating the said equation in the rainbow approximation. Working in Landau gauge and neglecting wavefunction renormalization effects. The resulting gap equation has a similar form as in the unquenched case of~\cite{Gorbar:2001qt}, but the effective coupling has a rather different physical interpretation, as it corresponds to the bare electromagnetic coupling.

 This scenario has been also considered at finite temperature by the same group~\cite{Nascimento:2015ola}. Provided the coupling exceeds the critical value {\em in vacuum}, these authors  estimate the ratio of the dynamical mass in vacuum and the critical temperature to be of order $2\pi$. In this article, we revisit the  calculations in~\cite{Gorbar:2001qt,Alves:2013bna,Nascimento:2015ola} within the static or constant mass approximation~\cite{Ahmad:2015cgh}. We first confirm the critical value for the effective coupling above which chiral symmetry is broken in vacuum. We then explore the behavior of the dynamical mass and critical temperature for symmetry restoration. 
We present our findings in the following manner. Section~\ref{vac} we describe the Lagrangian and Feynman rules of the theory. We also discuss the conditions for chiral symmetry breaking by solving the corresponding SDE. We promote this equation at finite temperature in Sect.~\ref{FT}. We introduce the so-called Constant Mass Approximation (CMA) in Sect.~\ref{CMA} and  conclude  in Sect.~\ref{conclu}

\section{Chiral Symmetry Breaking in vacuum}\label{vac}

Feynman rules for RQED follow from the Lagrangian
\be
{\cal L} = -\frac{1}{4}F^{\mu\nu} \frac{2}{(-\square)^{1/2}} F_{\mu\nu} +\bar{\psi}(i\gamma^\mu \partial_\mu+e\gamma^\mu A_\mu)\psi,\label{lagrangian}
\ee
where $F^{\mu\nu}$ is the electromagnetic field tensor and $A^\mu$ the corresponding gauge field, $e$ is the electric charge and $\psi$ the 4-component fermion field. Dirac matrices are represented by the $4\times 4$  matrices $\gamma^\mu$. Greek indices run from 0 to 2 and $(-\square)$ is the corresponding D'\ Alambertian operator, which appears in the Lagrangian with fractional power.

The bare 2-point functions derived from~(\ref{lagrangian}) are~\cite{Marino:1992xi,Gorbar:2001qt,Teber:2012de}
\be
\Delta_{\mu\nu}^{(0)}(q)= \frac{1}{2q}\left(\delta_{\mu\nu}-\frac{q_\mu q_\nu}{q^2} \right),
\label{photon}
\ee
which corresponds to the Landau-like gauge bare {\em photon} propagator. Notice the softening of the infrared divergence of the propagator, resulting from the non-perturbative integration of the third component of the stress tensor of the ordinary electromagnetic field coupled to the matter field in the plane of motion of electrons. The massless fermion propagator remains
\be
S_0^{-1}(p)=-\gamma^\mu p_\mu.
\ee
 Non-perturbatively, the Schwinger-Dyson equation (SDE) for the latter is expressed as
\be
S^{-1}(p)=S_0^{-1}(p) - \Xi(p),\label{SDE}
\ee
where the {\em electron} self-energy is
\be
\Xi(p)=-e^2 \int\frac{d^3k}{(2\pi)^3} \gamma^\mu S(k) \Gamma^\nu(k,p) \Delta_{\mu\nu}(q),\label{self}
\ee
where $q=k-p$ and $\Gamma^\nu(k,p)$ and $\Delta_{\mu\nu}(q)$ represent to the full electron-photon vertex and full photon propagator. Given a particular form of these Green functions, the general solution to Eq.~(\ref{SDE}) is
\be
S^{-1}(p) = -A(p) \gamma^\mu p_\mu +\Sigma(p).\label{sol}
\ee
Following the conventions of Refs.~\cite{Alves:2013bna,Nascimento:2015ola}, we search for the corresponding solution within the so-called rainbow-ladder truncation, where we replace $\Gamma^\nu(k,p)\to\gamma^\nu$ and $\Delta_{\mu\nu}(q)\to\Delta_{\mu\nu}^{(0)}(q)$ given in~(\ref{photon}). As a further simplification, we neglect wavefunction renormalization effects by setting $A(p)=1$, such that from (\ref{self}), the mass function $\Sigma(p)$ verifies the gap equation
\begin{equation}
    \Sigma(p) =4\pi\alpha
    \int \frac{d^3k}{(2\pi)^3}
    \frac{\Sigma(k)}{k^2+\Sigma(k)^2}
\frac{1}{\mid p-k \mid},\label{gap}
\end{equation}
where $\alpha=e^2/(4\pi)$.
It has been discussed in Ref.~\cite{Alves:2013bna} that in order to have a non-trivial solution to the above Eq.~(\ref{gap}), which would correspond to a chiral symmetry breaking solution, the coupling should exceed the critical value $\alpha_c=\pi/8$.  Moreover, the dynamical mass follows a Miransky scaling law. This expression is identical to Eq.~(22) of Ref.~\cite{Gorbar:2001qt} if we identify
\be
\pi \lambda = \alpha \qquad \lambda= \frac{e^2}{2\pi^2\left( 1+\frac{n_f e^2}{16}\right)}.\label{couplinggus}
\ee
In the remaining of this article we review this scenario in a hot medium characterized by a temperature $T$.


\section{SDE at finite T}\label{FT}

At finite temperature, within the Matsubara formalism, we replace any integral over the temporal component of any  four-vector $p=(p_0,{\bf p})$ by the summation 
\begin{equation}
    \int dp_0f(p_0) \to T \sum_{n=-\infty}^\infty f(i\omega_n),
\end{equation}
where the fermionic Matsubara frequencies are $\omega_n=(2n+1)\pi T$.
In this formalism, the gap equation reads~\cite{Nascimento:2015ola}
\begin{equation}
    \Sigma_m({\bf p}) =
    \sum_{n=-\infty}^{\infty} \int_\Lambda \frac{d^2k}{(2\pi)^2}
    \frac{(4\pi\alpha T) \Sigma_n({\bf k})}{(2n+1)^2\pi^2T^2+{\bf k}^2+\Sigma_n({\bf k})^2}
\frac{1}{[4 (m-n)^2\pi^2T^2+({\bf p}-{\bf k})^2 ]^{1/2}},
\end{equation}
where we have used the shorthand notation $\Sigma_n({\bf p})=\Sigma(\omega_n,{\bf p})$ and the symbol $\int_\Lambda$ refers to the fact that divergent integrals are to be regularized with an ultraviolet cut-off $\Lambda$.
%
%
%

\begin{figure}
 \centering
    \includegraphics[width=0.45\textwidth]{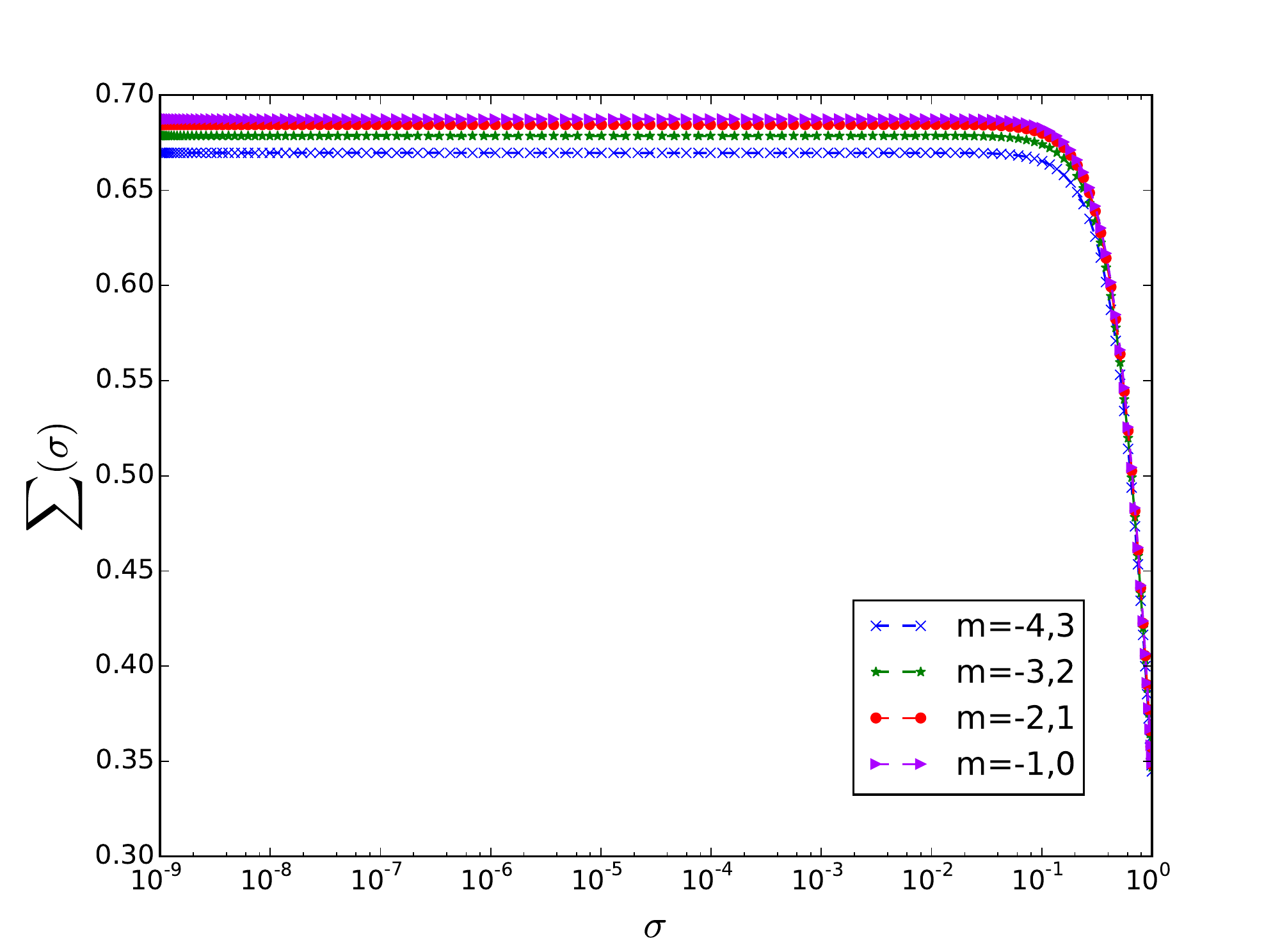}
    \includegraphics[width=0.45\textwidth]{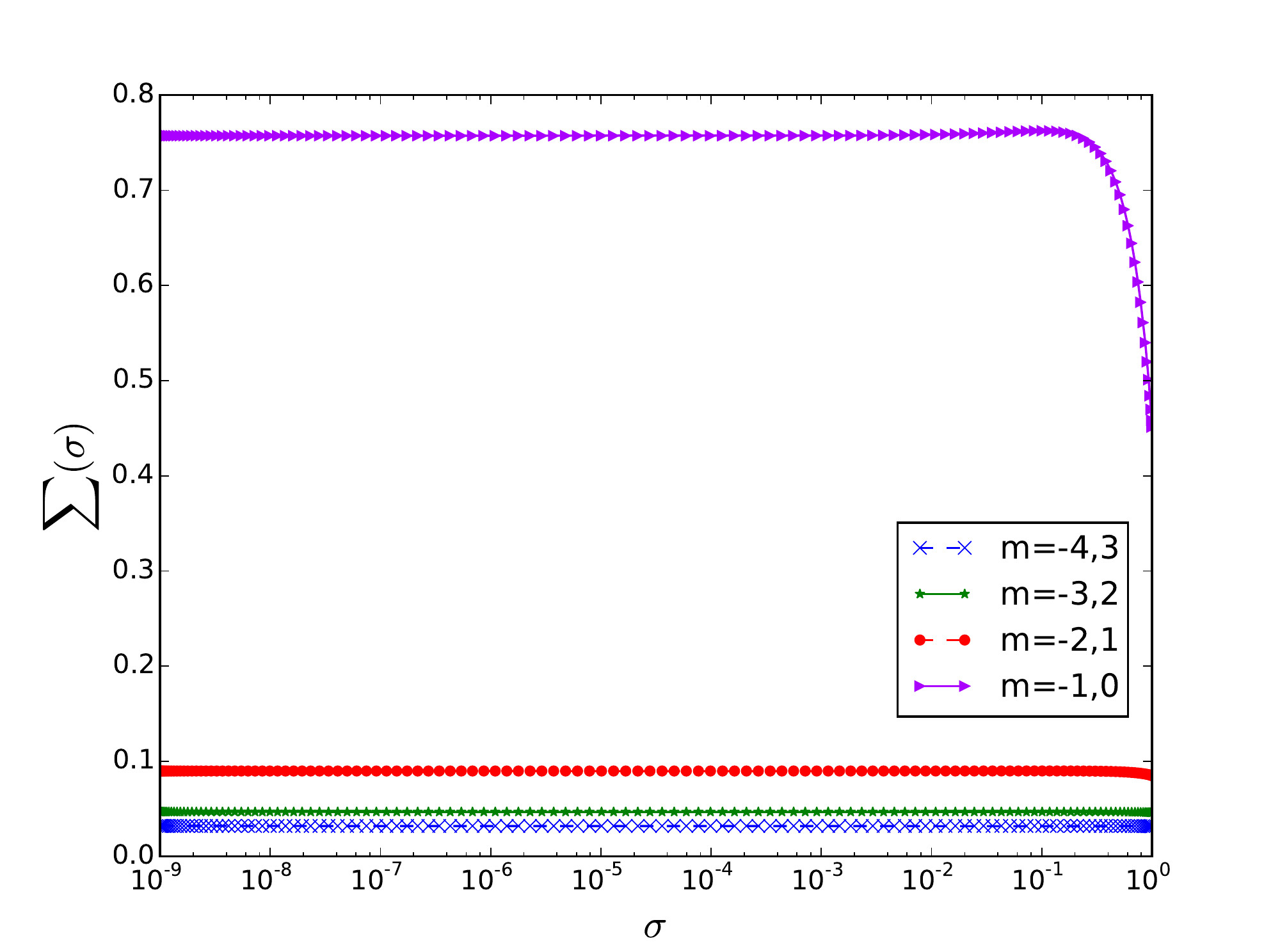}
    \caption{Mass function as a function of $\sigma$ for  $\alpha=14\alpha_c$ with $N_{f}=7$ Matsubara frequencies. {\em Left panel:}  $\tilde{T}$=0.005; {\em right panel}, $\tilde{T}$=0.95.  Symbols correspond to different Matsubara numbers for which the functions $\Sigma_m(\sigma)$  are identically equal.}
    \label{fig1}
\end{figure}

Introducing the dimensionless quantities,
\begin{equation}
\mathbf{p}=\Lambda\boldsymbol{\sigma},\quad \mathbf{k}=\Lambda\boldsymbol{\rho},\quad
T=\Lambda \Tilde{T},\quad
\Sigma_m(\mathbf{k})=\Lambda \tilde{\Sigma}_m(\boldsymbol{\rho}),
\end{equation}
the gap equation becomes
\be
\tilde{\Sigma}_m (\mathbf{\sigma})  =
\alpha \Tilde{T}
\sum_{n=-N_f-1}^{N_f} \int_0^{1}
\frac{d\rho}{\pi} \rho d\theta
 \frac{ \tilde{\Sigma}_n(\mathbf{\rho})}{(2n+1)^2\pi^2 \tilde{T}^2+ \rho^2+\tilde{\Sigma}_n^2(\mathbf{\rho})}
\frac{1}{\left[4 (m-n)^2\pi^2\tilde{T}^2+ ({\boldsymbol\sigma}-{\boldsymbol\rho})^2  \right]^{1/2}},\label{gapdimless}
\ee
where $\theta$ is the angle between the vectors $\boldsymbol{\sigma}$, $\boldsymbol{\rho}$ with magnitudes $\rho=|\boldsymbol\rho|$ and $\sigma=|\boldsymbol{\sigma}|$, respectively. In this expression, the cut-off does not appear in the momentum integrations, but in the number of Matsubara frequencies $N_f$ that are summed up.
 We solve the above equation~(\ref{gapdimless})  fixing $\alpha$ and $T$ appropriately and then recursively search for the solution starting from a given numerical seed. The double numerical integration is performed using Gaussian quadratures, after rescaling the radial component of momentum. In Fig.~\ref{fig1}  we show, for the sake of illustration, the solution of the gap equation as a function of the momentum with $N_f=7$ Matsubara frequencies at two different temperatures and the same fixed value of $\alpha>\alpha_c$. We observe that when $T$ is close to zero (left panel), every mass function $\Sigma_m(\sigma)$ has almost the same height as $\sigma\to 0$, but the larger the temperature, the height of all $\Sigma_m(\sigma)$ diminishes, except for the corresponding to $m=0$ (right panel).

 Once we are able to identify a stable solution for each $\tilde{T}$,  we explore the behavior of the chiral condensate, which is the order parameter for the chiral transition and is defined as 
 
 \bea
\langle\bar{\psi} \psi\rangle&=&\operatorname{Tr} i S(0)\nonumber\\
&=&\frac{\tilde{T}}{2 \pi} \sum_{m=-N_f-1}^{N_f} \int_0^1 d\rho \rho
\frac{ \tilde{\Sigma}_{m}(\rho)}{(2 m+1)^{2} \pi^{2} \tilde{T}^{2}+\rho^{2}+\tilde{\Sigma}_{m}^{2}(\rho)}.
\eea
This condensate is finite when chiral symmetry is broken, and vanishes when this symmetry is restored. The value of $\tilde{T}$ at which this happens is $\tilde{T}_c$, the critical temperature for the chiral symmetry restoration.


\section{Chiral symmetry breaking at finite $T$ and constant mass approximation}\label{CMA}

\begin{figure}[h]
    \begin{center}
    \includegraphics[width=0.6\textwidth]{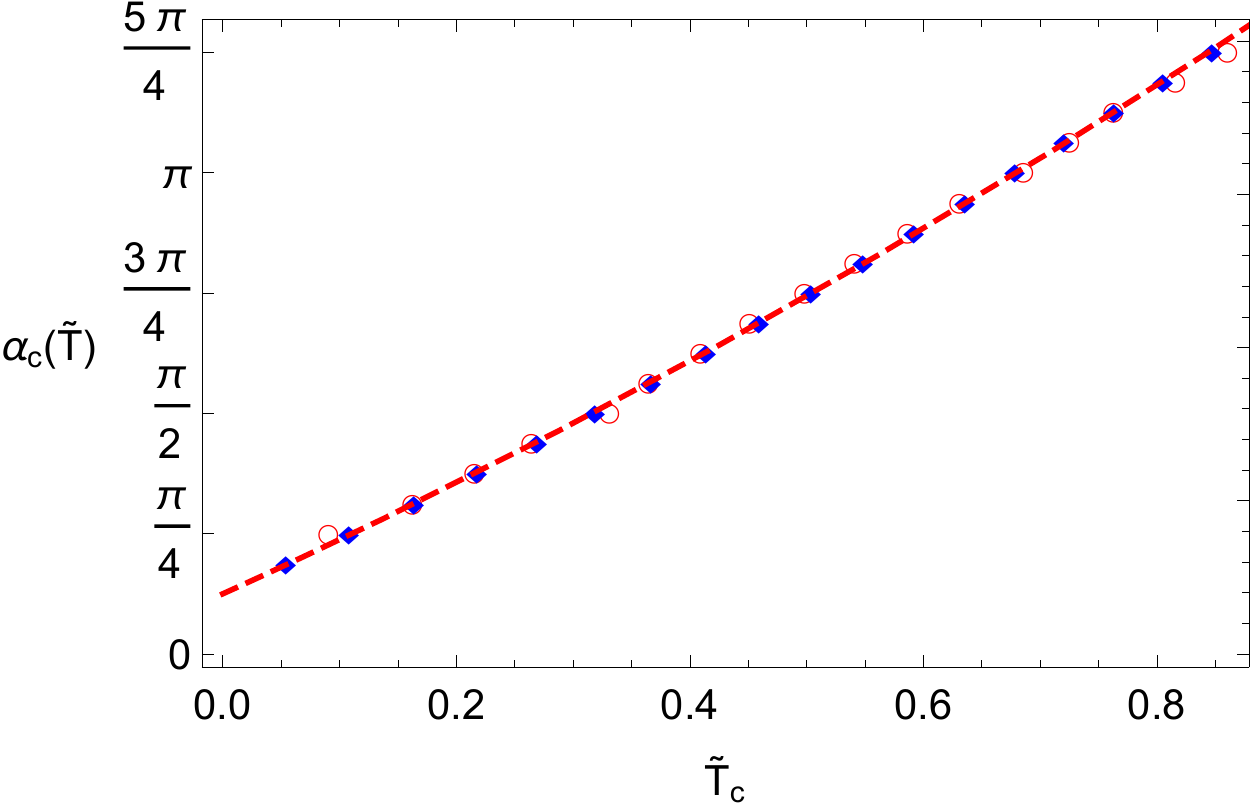}
    \end{center}
    \caption{$\alpha_c$ as a function of the critical temperature summing up to $N_f=7$. Blue diamonds  correspond to eq.~(\ref{ac}), whereas open red circles correspond to the values extracted from the vanishing of the chiral condensate in the constant mass approximation, eq.~(\ref{condcma}).
    }
    \label{ac_vs_T}
\end{figure}

Because chiral symmetry breaking is an infrared phenomenon, it basically is encoded in the behavior of $\Sigma_m(0)$. Therefore, 
in the constant mass approximation~\cite{Ahmad:2015cgh}, we replace all mass functions involved in the SDE with their values at zero momentum. Denoting $\tilde{\Sigma}_m({\bf{\sigma}})\rightarrow S_m$, the gap equation becomes
\be
S_m =
2\alpha \Tilde{T}
\sum_{n=-N_f-1}^{N_f} \int_0^{1}
d\rho \rho
    \frac{ S_n}{(2n+1)^2\pi^2 \tilde{T}^2+ \rho^2+S_n^2}
\frac{1}{\left[4 (m-n)^2\pi^2\tilde{T}^2+ \rho^2  \right]^{1/2}}.
\ee
In what follows, we select  the cut-off $N_f$ such that 
\be
(2N_f+1)\pi T_0 =\Lambda \Rightarrow (2N_f+1)\pi \tilde{T}_0 =1
\rightarrow \tilde{T}_0=\frac{1}{(2N_f+1)\pi},
\ee
and write all temperature scales proportional to multiples of $ \tilde{T} _0 $, namely,
$T = k \tilde {T} _0,\;k\in \mathbb{N} $.
Performing momentum integration analytically, we have that
\bea
S_m  &=&
2\alpha \Tilde{T} \sum_{n=-N_f-1}^{N_f}
\frac{S_n}{\sqrt{\pi ^2
   \Tilde{T}^2 \left(-4 m^2+8 m n+4 n+1\right)+S_n^2}} 
\nonumber \\
&\times& 
\left[\tan^{-1}
\sqrt{
\frac{4 \pi ^2 \Tilde{T}^2 (m-n)^2+1}
{\pi^2 \Tilde{T}^2 \left(-4 m^2+8 m n+4 n+1\right)+S_n^2}
   }\right.
   \nonumber \\
&&
\left. 
   -\tan ^{-1} \frac{2 \pi 
   \Tilde{T} \left| m-n\right| }{\sqrt{\pi ^2 \Tilde{T}^2 \left(-4 m^2+8
   m n+4 n+1\right)+S_n^2}} \right].
   \label{sdmc}
\end{eqnarray}
This transcendental equation can be solved self consistently, giving the behavior of each $S_m$ as a function of temperature. 
Near the critical point, we expect $ S_m $ to approach zero.
Assuming a linear behavior of $ S_m \approx \gamma_m (T-T_c) $ near criticality, we have that
\begin{eqnarray}
S_m  &\approx&
2\alpha \Tilde{T} \sum_{n=-N_f-1}^{N_f}
\frac{S_n}{\sqrt{\pi ^2
   \Tilde{T}^2 \left(-4 m^2+8 m n+4 n+1\right)}} 
\nonumber
   \\
&&\hspace{-8mm}\times
\left[\tan ^{-1}
\sqrt{
\frac{4 \pi ^2 \Tilde{T}^2 (m-n)^2+1}
{\pi^2 \Tilde{T}^2 \left(-4 m^2+8 m n+4 n+1\right)}
   } 
 -  \tan ^{-1} \frac{2  \left| m-n\right| }{\sqrt{ \left(-4 m^2+8
   m n+4 n+1\right)}} \right].
\end{eqnarray}
For the zeroth Matsubara frequency $m=0$, the above relation simplifies to 
\begin{eqnarray}
\frac{1}{2\alpha_c \Tilde{T}_c}  &=&
 \sum_{n=-N_f-1}^{N_f}
\frac{S_n/S_0}{ \pi \Tilde{T}_c \sqrt{ \left(4 n+1\right)}} 
\left[\tan ^{-1}
\sqrt{
\frac{4 \pi^2 \Tilde{T}^2_c n^2+1}
{\pi^2 \Tilde{T}^2_c \left(4 n+1\right)}
   } 
   -\tan ^{-1} \frac{2 \left| n \right| }{\sqrt{\left(4 n+1\right)}} \right].
\end{eqnarray}
Now, because $S_n/S_0\leq 1$, it follows that
\begin{eqnarray}
\frac{1}{\alpha_c }  &=&
 \sum_{n=-N_f-1}^{N_f}
\frac{2 S_n/S_0}{ \pi  \sqrt{ \left(4 n+1\right)}} 
\left[\tan ^{-1}
\sqrt{
\frac{4 \pi^2 \Tilde{T}^2_c n^2+1}
{\pi^2 \Tilde{T}^2_c \left(4 n+1\right)}
   } 
   -\tan ^{-1} \frac{2
    \left| n \right| }{\sqrt{ \left(4 n+1\right)}} \right].
\nonumber \\   
   &\leq&
 \sum_n
\frac{2}{ \pi  \sqrt{ \left(4 n+1\right)}} 
\left[\tan ^{-1}
\sqrt{
\frac{4 \pi^2 \Tilde{T}^2_c n^2+1}
{\pi^2 \Tilde{T}^2_c \left(4 n+1\right)}
   } 
   -\tan ^{-1} \frac{2 \left| n \right| }{\sqrt{\left(4 n+1\right)}} \right].
\end{eqnarray}
Because of our assumption that near the critical point the temperature-dependent $S_m(\tilde{T}) $ approaches to zero point perpendicularly, in consistency with the Clausius-Clapeyron criterion, we reach to the equality ($S_n/S_0\rightarrow 1 $):
\begin{eqnarray}
\frac{1}{\alpha_c } 
&=&
 \sum_{n=-N_f-1}^{N_f}
\frac{2}{ \pi  \sqrt{ \left(4 n+1\right)}} 
\left[\tan ^{-1}
\sqrt{
\frac{4 \pi^2 \Tilde{T}^2_c n^2+1}
{\pi^2 \Tilde{T}^2_c \left(4 n+1\right)}
   } 
   -\tan ^{-1} \frac{2 \left| n \right| }{\sqrt{\left(4 n+1\right)}} \right].\label{ac}
\end{eqnarray}
Summation over $n$ is finite for every value of temperature, which allow us to obtain the behavior of the critical coupling for each $\tilde{T}$.  Considering a cutoff $ N_f = 7 $, the behavior of the critical coupling as a function of $\tilde{T}$ can be observed in Fig.~\ref{ac_vs_T}.


The critical coupling $\alpha_c$ can also be obtained from the chiral condensate, which in this approximations reads

\begin{eqnarray}
\langle \bar\psi\psi\rangle &=&
\frac{\tilde{T}}{2 \pi} \sum_{m=-N_f-1}^{N_f}  
\int_0^1 d\rho \rho
\frac{ S_{m}}{(2 m+1)^{2} \pi^{2} \tilde{T}^{2}
+\rho^{2}+S_{m}^{2}}.\nonumber \\
&=&
\frac{\tilde{T}}{4 \pi}
\sum_{m=-N_f-1}^{N_f} 
\ln\left(1+ \frac{1}{(2 m+1)^{2} \pi^{2} \tilde{T}^{2}+S^2_m}
\right) \\
\label{condcma}
\end{eqnarray}
In the same Fig.~\ref{ac_vs_T}, a comparison is shown between the values of $\alpha_c$ derived from Eqs.~(\ref{ac}) and (\ref{condcma}), with a qualitative agreement. A  fit to this behavior is of the form 

\begin{equation}
\alpha_c (\tilde{T}) = \frac{\pi}{8} + 3.44 \tilde{T}+0.88 {\tilde{T}}^2,
\label{fit:ac}
\end{equation}

This behavior is in disagreement with the finding of  Ref.~\cite{Nascimento:2015ola}, which establishes that
\begin{equation}
\alpha_c({\tilde{T}})=\frac{1}{1-2\tilde{T}}.
\end{equation}
As $\tilde{T} \to 0$, the last expression is inconsistent with the zero temperature value of the critical coupling $\alpha_c=\pi/8$.

\begin{figure}[h]
    \centering
    \includegraphics[width=0.7\textwidth]{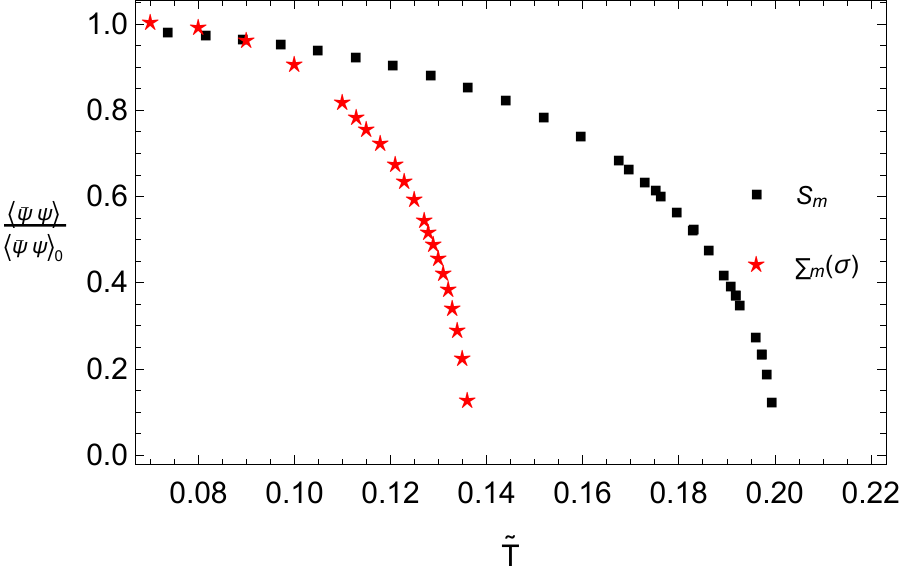}
    \caption{Chiral condensate $\langle \bar{\psi} \psi \rangle$ as a function of  $\tilde{T}$ for $\alpha= 3.0\alpha_c$ with $N_{f}=7$. Black squares correspond to the CMA, whereas red stars to the full momentum dependent mass function.}
    \label{cond_vs_T}
\end{figure}

Next, maintaining the fixed value of $ N_f =7$ and taking 
$\alpha= 3.0\alpha_c $, 
in Fig.~\ref{cond_vs_T} we show the behavior of the normalized condensate as a function of $\tilde{T}$. We also compare the findings of the condensate including the full momentum dependence of the mass functions. We observe that the number of Matsubara frequencies taken into account is enough to reproduce the correct physical behavior expected for the condensate, namely, it approaches the vertical axis roughly as a constant at small temperature, whereas it hits the horizontal axis with a vertical line as approaches to the critical temperature. The difference of the $T_c$ for both the approximations is due to the static nature of the CMA.



\begin{figure}[!h]
    \centering
    \includegraphics[width=0.7\textwidth]{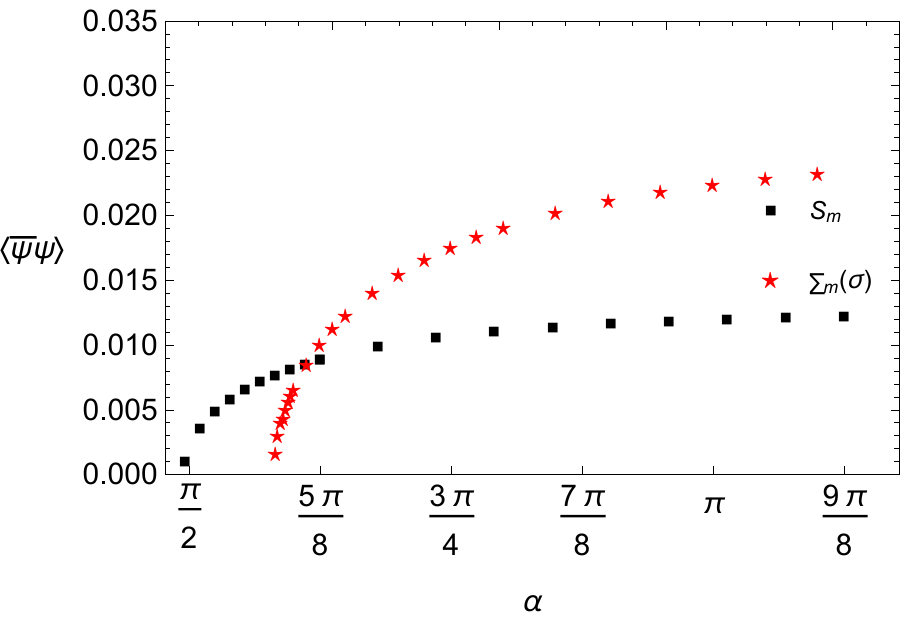}
    \caption{Chiral condensate as a function of the coupling for fixed $\tilde{T}=0.3$ with $N_{f}=7$. Black squares correspond to the CMA, whereas red stars to the full momentum dependent mass function. }
    \label{fig:my_label}
\end{figure}

Moreover, fixing the value of the temperature, the condensate as a function of the coupling is depicted in Fig.~\ref{fig:my_label}. Results from SDE and CMA are in qualitative agreement, namely, in both cases the condensate starts rising just above $\alpha_c$ and saturate at large $\alpha$. The difference of the $\alpha_c$ for this to happen is due to the nature of CMA.

\begin{figure}[!h]
    \centering
    \includegraphics[width=0.7\textwidth]{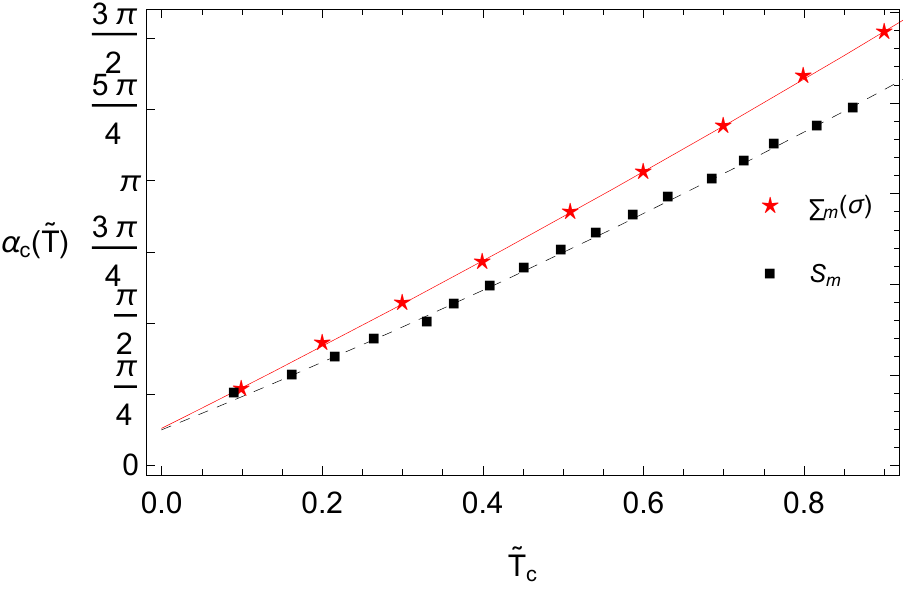}
    \caption{$\alpha_c$ as a function of $\tilde{T}_c$ with $N_{f}=7$ fixed.  Red stars correspond to the behavior derived from $\Sigma_{n}(p)$ and black squares to $\Sigma_{n}$. Continuous curves are quadratic fits of Eqs.~(\ref{quadfitCMA}) (dashed) and~(\ref{quadfit}) (solid).  }
    \label{ac_vs_T_s(p)}
\end{figure}

We also depict the critical coupling as a function of $\tilde{T}_c$ from these condensates in Fig.~\ref{ac_vs_T_s(p)}. The behavior is qualitatively the same and the error of the CMA does not exceed a few percent in the range of temperatures of our framework as compare with the full SDE prediction. The behavior remains quadratic, of the form

\begin{equation}
\alpha_c=\frac{\pi}{8} + 3.58 \tilde{T} + 0.649 \tilde{T}^{2},
\label{quadfitCMA}
\end{equation}
for the CMA (dashed curve) and 
\be
\alpha_c=\frac{\pi}{8} + 4.42 \tilde{T} + 0.487 \tilde{T}^{2},
\label{quadfit}
\ee
for $\Sigma_{n}(p)$ (solid curve). 

\begin{figure}[h]
    \centering
    \includegraphics[width=0.7\textwidth]{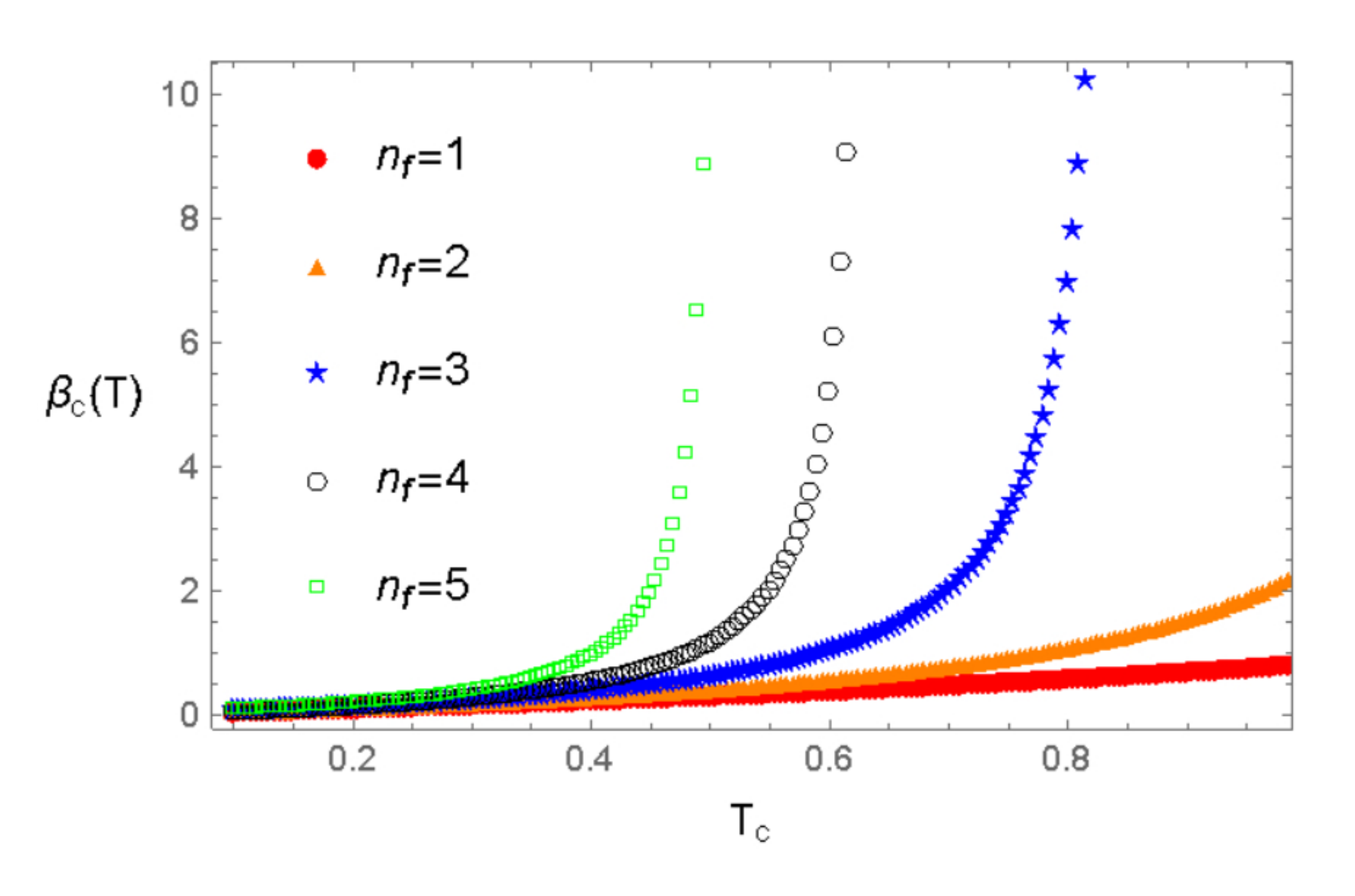}
    \caption{Singular behavior for effective coupling $\beta_c$ taking several number of families $n_f$.}
    \label{bc_vs_Tc}
\end{figure}

If we consider vacuum polarization effect \cite{Gorbar:2001qt}, we show the effective coupling as a function of temperature for different fermion families $n_f$
in figure \ref{bc_vs_Tc}.

\begin{equation}
    \beta_{c}=\frac{\alpha}{2-\alpha(\frac{\pi n_{f}}{4})}
\end{equation}



 
\section{Conclusions}\label{conclu}


 In this article, we have revisited the behavior of chiral symmetry restoration in RQED in a heat bath. First, we have computed the critical coupling for chiral symmetry breaking by solving the SDE  numerically.  Then, the critical temperature is obtained at the point in parameter space in the supercritical regime where the chiral condensate vanishes. We have further approximated the solution to the gap equation in the so-called constant mass approximation by neglecting any momentum dependence of the mass function and taking into account only their IR value. Neither of these numerical solutions  agree with the behavior predicted by \cite{Nascimento:2015ola}. Our numerical procedure shows that there is no critical behavior for none of the values of the cutoff. Instead, the behavior of the critical coupling behaves as a second order polynomial of Temperature. Furthermore, in the limit $T=0$, the predicted critical coupling for the momentum dependent mass coincides with the critical coupling predicted within the CMA. Extensions to this work are currently under consideration by adding vacuum polarization effects and a chemical potential. Results shall be reported elsewhere.

\begin{acknowledgements}
The authors would like to thank Valery P. Gusynin for valuable comments.
JCR acknowledges support
from FONDECYT (Chile) under Grant No. 1170107.

\end{acknowledgements}

\end{document}